# Chemical Abundances in Broad Emission Line Regions: The "Nitrogen-Loud" QSO 0353-383

J.A. Baldwin

*Physics and Astronomy Department, Michigan State University, 3270 Biomedical Physical Sciences Building, East Lansing, MI 48824*

F. Hamann

*Department of Astronomy, University of Florida, 211 Bryant Space Center, Gainesville, FL 32611-2055*

K.T. Korista

*Department of Physics, University of Western Michigan, 1120 Everett Tower, Kalamazoo, MI 49008-5252*

G.J. Ferland

*Department of Physics and Astronomy, University of Kentucky, 177 Chemistry/Physics Building, Lexington, KY 40506*

M. Dietrich and C. Warner

*Department of Astronomy, University of Florida, 211 Bryant Space Center, Gainesville, FL 32611-2055*

baldwin@pa.msu.edu

## Abstract

The intensity of the strong N V $\lambda$1240 line relative to C IV $\lambda$1549 or to He II $\lambda$1640 has been proposed as an indicator of the metallicity of QSO broad emission line regions, allowing abundance measurements in a large number of QSOs out to the highest redshifts. Previously, it had been shown that the (normally) much weaker lines N III] $\lambda$1750 and N IV] $\lambda$1486 could be used in the same way. The redshift 1.96 QSO 0353-383 has long been known to have N III] and N IV] lines that are far stronger relative to Ly$\alpha$ or C IV than in any other QSO. Because in this particular case these intercombination lines can be easily measured, this unusual object provides an ideal opportunity for testing whether the N V line is a valid abundance indicator. Using new observations of Q0353-383 made both with HST in the ultraviolet and from the ground in the visible passband, we have carefully remeasured the emission lines and reanalyzed their strengths using the latest models of the QSO broad emission line region. We find that intensity ratios involving the strengths of N V, N IV] and N III] relative to lines of He, C and O all indicate that nitrogen is overabundant relative to oxygen in Q0353-383 by a factor of ~15 compared to solar ratios. This agreement among the diagnostics supports the use of these lines for measuring BLR chemical abundances. If nitrogen behaves like a secondary element, such that N/O $\propto$ O/H, then the extreme nitrogen enhancement in Q0353-383 implies a metallicity of ~15 times the solar value. Even if Q0353-383 represents an extreme outlier in the N/O ∝ O/H relation, the overall metallicity should still be *at least* five times solar. Unusually high metallicities in Q0353-383 might imply that we caught this object just as the gas-phase metallicity in the central part of its host galaxy has peaked, at a time when the interstellar gas supply is nearly exhausted and hence the fuel source for the central QSO is ready to shut off.

*Subject headings:* quasars: emission lines, galaxies: active

# 1.0 Introduction

The quasi-stellar object Q0353-383 was discovered by Osmer & Smith (1980) and found to have a redshift $z = 1.96$. It is a luminous object, with a continuum level only about a factor of 10 lower than those of the most luminous known QSOs.

It was quickly seen that this QSO has a very unusual emission-line spectrum, in which the normal ultraviolet lines are present but are supplemented by exceptionally strong N III] λ1750 and N IV] λ1486 intercombination lines. Osmer (1980) analyzed this spectrum using the best observations and models possible at the time and concluded that this object probably has an anomalously high nitrogen abundance.

In the intervening years it has been realized that the abundance of nitrogen relative to the other heavy elements (C,O, etc.) is a good marker of the degree of chemical enrichment of the interstellar medium in galaxies. This is because N is built up as a secondary element in the CNO cycle, so that at higher metallicities N/H ∝ $(O/H)^2$, or N/O ∝ O/H ∝ $Z$, where $Z$ is defined as $Z/Z_{sun}$ ~ (metals/H)/(metals/H)$_{sun}$ ~ (O/H)/(O/H)$_{sun}$ (see Van Zee et al. 1998, Pettini et al. 2002 and the discussion in Hamann et al. 2002, Appendix A) . Shields (1976) introduced the idea of measuring the relative C, N and O abundances in the BLR, using intercombination lines including the N III] and N IV] lines which are usually quite weak, but which are strong in Q0353-383. Hamann & Ferland (1993) showed that the strong N V emission line also can be used (see also Hamann & Ferland 1999 and references therein). In QSOs with the luminosity of Q0353-383, the mass of the ionized gas in the BLR must be at least $10^3$–$10^4$ M$_{sun}$ (Baldwin et al. 2002). Assuming that this BLR gas is being accreted from the surrounding galaxy, it must be averaging over the properties of a bulge-sized stellar system. Therefore, these methods for determining chemical abundances in the BLR offer a way to trace the history of metal enrichment in the centers of (proto)galaxies at large lookback times.

An improvement in the models since Osmer's study is the realization that the spectrum emitted by the BLR is not nearly as sensitive to the (unknown) structure of the BLR as had been feared (Baldwin et al. 1995). Reverberation studies (Peterson 1993) show that the BLR gas is widely distributed, and once that is the case the total emitted spectrum is much more sensitive to the chemical abundances and the shape of the ionizing continuum than it is to the details of the BLR structure (cf. Korista et al. 1998).

Therefore, a realistic long-term goal in the study of QSO emission-line spectra is to use the relative strengths of their nitrogen lines to trace the early chemical evolution of (proto)galaxies. Q0353-383, with its abnormally strong N III] and N IV] lines, must be an extreme object of one sort or another, and therefore offers an important test of whether or not the nitrogen line strengths really are reliable abundance indicators.

Here we describe a new analysis of Q0353-383, using new spectra and up-to-date models. We obtained higher-quality ground-based spectra than were previously available, and we used HST to obtain ultraviolet spectra in order to measure strengths or upper limits for the many emission lines expected at wavelengths below that of Ly$\alpha$. Our goal is to determine whether the emission-line spectrum of Q0353-383 really does indicate exceptionally high metal abundances, or whether some other mechanism is responsible for the unusual strengths of the nitrogen lines.

## 2.0 Observations

### 2.1 Ground-based spectroscopy

We used the RC spectrograph on the Blanco 4m telescope at CTIO to obtain ~200 km s$^{-1}$ resolution spectra of Q0353-383 on 18 Oct 1999 UT. The 3050-5750Å blue region (rest $\lambda\lambda$1030–1943Å) was observed using grating KPGL1, which gave 3 Å FWHM resolution through a 1.5 arcsec slit. The $\lambda\lambda$4860-9900 Å red region (rest $\lambda\lambda$1642–3345Å) was observed on the same night using grating G181 and a GG495 order-separator filter, with 6 Å resolution. The night was photometric, and for each setup short exposures of Q0353-383 and of three standard stars (from Hamuy et al 1994) were obtained through a 6 arcsec slit in order to determine accurate flux calibrations for the much longer exposures of Q0353-383 taken through the 1.5 arcsec slit. The total exposure times for these latter spectra were 8400 s in the blue and 2700 s in the red. Several separate exposures using the narrow slit were median-filtered together in order to remove cosmic rays. The data were reduced using standard IRAF software.

The reduction procedure involved extending the flux calibration as far into the blue as possible in order to join the ground-based spectrum to the end of the HST spectrum at $\lambda$3066 Å (rest $\lambda$1036Å). The flux measurements of the standard stars given by Hamuy et al. stop at $\lambda$3300 Å. The earth's atmospheric transmission strongly cuts off between this wavelength and $\lambda$3050 Å, so this is the area of the spectrum where we most need a high density of calibration points in order to properly determine the wavelength dependence of the correction for this absorption. Our approach was to extrapolate the spectrum of one of the Hamuy et al standards, H600, down to $\lambda$3050 Å by fitting it to stars of similar spectral types from the Bruzual-Persson-Gunn-Stryker Spectrophotometry Atlas (Bruzual et al. 1996), which are calibrated down to 1150 Å. We found that the best extrapolation was to assume that the spectrum of H600 is flat in $F_\lambda$ between $\lambda$3300 and $\lambda$3000 Å . Our flux calibration used only this extrapolation in the region below $\lambda$3300 Å, but the Hamuy et al. calibrations of all three standards (H600, LTT 1788 and LTT 2415) were used from $\lambda$3300 Å redwards.

The red spectra also had to be corrected for weak (5% peak-to-peak) interference fringes in the wavelength range to the red of $\lambda$7500Å. The fringes were calibrated from the spectrum of a standard star. There is excellent photometric agreement between the overlapping parts of the blue and red spectra. The two overlap between $\lambda\lambda$4860-5750 Å, and their flux levels agree to within the width of the noise.

## 2.2 HST spectroscopy

STIS on the Hubble Space Telescope was used to obtain a low-resolution ultra-violet spectrum covering the observed wavelength range $\lambda\lambda$1684-3066 Å (rest $\lambda\lambda$569–1036Å). We used the CCD detector, grating G230LB and the 52×0.2 aperture, which produce ~2.7 Å (330 km s$^{-1}$) FWHM resolution. A total of 7940 s of data were taken on 28 Jan 2000 UT as a series of 3 exposures, each with CRSPLIT = 2. The spectra were reduced through the standard STIS pipeline, and then median-filtered together to remove cosmic rays.

Q0353-383 had not previously been observed in the ultraviolet, so we did not know whether or not there would be a cutoff due to Ly-continuum absorption, or to what extent Ly$\alpha$ forest absorption lines would blanket specific emission lines of interest. Our HST spectra therefore were designed to be a reconnaissance. The instrumental setup was chosen as a compromise between obtaining reasonable sensitivity and wavelength coverage, and having enough spectral resolution to get some idea of the shape of the background QSO spectrum between the individual Ly$\alpha$ forest absorption lines.

The combined spectrum is plotted on Figure 1, with heavy lines (the light lines show the composite spectrum described in §3.6). The HST spectrum is joined to the blue ground-based spectrum at $\lambda$3060 Å, with no rescaling of the absolute flux in either spectrum. Figure 2 shows the overlapping portions of these two spectra before they were combined, illustrating the excellent photometric agreement between them. We deduce that our blue flux calibration is accurate down to about $\lambda$3050 Å, which is important because the strong OVI 1035 emission line straddles the joint between the HST and blue spectra.

The HST data were taken 3 months later than the ground-based data, and so we have to worry about potential variability of either the continuum or the emission-line spectrum. Typical variability timescales for luminous QSOs are in the range of many months to years (Kaspi et al. 2000), considerably longer and with lower amplitudes than for Seyfert galaxies. Figure 2 shows that the continuum matches up very well across the joint at 3050Å, implying that there was no continuum variation. However, this does not rule out variations in the emission lines, because the BLR is probably a light-year or more across and thus may require 1 or more years to respond to continuum variations. We compared our new ground-based spectra to an unpublished spectrum that we took at CTIO in November 1992, and found that the emission line and continuum fluxes all agree to within 2-3%. We also compared our data to the spectrum that Osmer took in 1976. Although our measured line fluxes and equivalent widths are in some cases rather different than those given by Osmer, we find by overplotting the two spectra that the agreement in the continuum shape and absolute flux is extremely good, and the emission lines agree to within the differences that might be expected from the very different instrumental resolutions that were used. We thus conclude that variability is not likely to be a problem.

## 3.0 Emission Line Measurements

The emission lines from Q0353-383 are narrow, which greatly facilitates the separation of blended lines. However, examination of the spectrum shows that the lines do not all have the same profiles. This is illustrated in Figure 3, where the upper five panels show the C IV 1549 doublet, Lyα λ1215 (blended with N V λ1240), N IV] λ1486, He II λ1640 (blended with O III] λ1663) and C III] λ1909 (blended with Si III] λ1892) all plotted on the same velocity system and scaled to have the same peak intensities.

The zero-point of the velocity scale in Fig. 3 is the average velocity of the peaks of the Lyα and C IV lines, at redshift $z = 1.960$. The C IV blend is plotted at rest wavelength λ1549.5 Å. The C IV doublet separation is 500 km s$^{-1}$, but it is still clear that the shape and peak velocity of C IV are in good agreement with Lyα, after disregarding the part of the Lyα profile to the red of +1000 km s$^{-1}$ where blending with N V starts to become significant. The peak of the He II profile lies 260 km s$^{-1}$ to the blue of the Lyα peak, while the N IV] peak is an additional 250 km s$^{-1}$ to the blue of He II. In addition, the wings of the C IV, N IV] and He II profiles have significantly different shapes. The core of the C III] profile is seen to have the same peak velocity as C IV and Lyα.

The strengths of the emission lines were measured by first isolating continuum-subtracted profiles of each line or blend of lines, then fitting template line profiles to these features in order to measure the line intensities. The continuum was fitted with linear or quadratic polynomials, using least-squares fits through all of the data points over a full range of a few hundred Å to either side of the emission line and that appeared to be continuum points rather than in emission or absorption lines.

At least three differently-shaped templates are required to be able to fit all of the different emission line profiles. The profiles of some individual lines can be described by just one of these templates, while other lines require a mixture of more than one template. The blends were modeled by adding together template profiles shifted to the positions of the individual lines in the blends, constraining their relative intensities in cases where the intensity ratios are fixed by the atomic physics. The remaining panels in Figure 3 show a deblended He II profile and two template profiles that will be described below.

## 3.1 The Lyα + N V blend.

The central core of the Lyα profile agrees remarkably well with the C IV λ1549 profile, and there is a well-defined feature at the expected position of the N V line. However, there is also a strong wing reaching redwards from the Lyα core and extending beneath the N V bump. It is unclear how much of this red wing is due to Lyα, and how much is due to N V or some other emission feature.

We fitted the blend as is shown in Figure 4a, assuming that both Lyα and NV consist of broad bases having the same profile as NIV], combined with a narrow core at approximately zero velocity. To find the shape of the narrow Lyα core, we fitted the N IV] profile to the wings of Lyα and to the two members (in a 3:2 intensity ratio) of the N V doublet. We then subtracted this fit from the observed blend. This residual spike has its first moment velocity at –74 km s$^{-1}$ and a full width at half maximum intensity FWHM = 1200 km s$^{-1}$. It is shown on Fig. 3 as the "narrow template". This may represent the narrow-lined region in Q0353-383.

The fit with the broad components also has a narrow bump left over at about λ1240 Å, which even though it cannot be fitted with the NV λλ1238, 1242 blend in any believable ratio, still can be interpreted as the narrow NV component. The sum of the narrow and broad components from this fit gives an upper limit to the flux in NV.

To estimate a lower limit to the NV strength, we cannot just consider the flux in the obvious bump between about $\lambda\lambda$1230 and 1245Å, because this feature clearly is too narrow to include the full NV profile. To make some sensible allowance for broad wings on NV, we fitted the "CIV template" profile described below in §3.2. We placed two such profiles in the expected 3:2 intensity ratio at the positions of the two N V components, assigning them as much flux as was possible without creating an obvious hole in the residual spectrum. The residual from this new N V fit includes the central part of Ly$\alpha$ plus a strong redward wing extending beneath N V, as is shown in Figure 4b. This wing could be an extra component seen only in Ly$\alpha$ because of the very much higher optical depth in this line. A mechanism such as Rayleigh scattering might lead to such a wing (cf. Korista & Ferland 1998).

These two methods of separating the Ly$\alpha$+N V blend gave us what we consider to be reasonable lower and upper limits to the actual N V flux. These are $1.5\times10^{-14} < F(\text{N V}) < 2.5\times10^{-14}$ erg cm$^{-2}$ s$^{-1}$, for the total N V doublet.

Clearly present at the extreme red edge of this blend is a weak line that we have labeled O I $\lambda$1305 in Figure 1. The feature is too broad to be just O I with any of the template profiles, so there may be some Si II $\lambda$1263 and $\lambda$1307 emission contributing as well. Si II $\lambda$1263 is reported by Laor et al. (1995) to be about 2–3% as strong as Ly$\alpha$ in a number of QSO spectra.

## 3.2 CIV 1549, CIII] 1909 and MgII 2798

We find that we can reconstruct the C IV $\lambda$1549 blend as a sum of the C IV $\lambda\lambda$1548.2, 1550.2 components in the intensity ratio 3:2 (consistent with the photoionized cloud models described below) using a mix of the N IV] and narrow template components. The best-fitting broad:narrow intensity ratio for the individual members of the C IV doublet is 1.08:1, and the total broad + narrow template profile reconstructed using this ratio is shown as the bottom panel in Fig. 3, labeled "C IV template".

The cores of the C III] and Mg II profiles are reasonably well fitted by the same C IV template profile. C III] $\lambda$1909 is blended with Si III] $\lambda$1892, which for the measurement in Table 1 was also fitted with a C IV template profile even though the fit is not perfect. The lines are narrow enough that Al III $\lambda$1857 would be well-separated from Si III], but it is not detected; the unidentified bump near this position is at too short a wavelength (its centroid lies 1800 km s$^{-1}$ to the red of the Al III position). There is weak excess emission on the red wing of C III], which we cannot identify. Mg II $\lambda$2798 is also badly blended on its wings, in this case with Fe II lines.

## 3.3 He II 1640

The fourth panel down in Figure 3 shows the observed He II $\lambda$1640.7 + OIII] $\lambda\lambda$1660.8, 1666.1 blend. Since there is only modest overlap in velocity between He II and O III], any sensible fit to O III] will leave behind a nearly correct He II profile. We tried to fit the O III] doublet with our template profiles because we were interested in trying to subtract its blue wing from the red wing of He II as accurately as possible. None of the templates provided really good fits; using the C IV template profile left behind excess flux on the blue wing of the O III] bump while using the (deblended) He II profiles left excess flux on the red wing. Although we can't really tell the true shape of the O III] lines, these different fits lead to an uncertainty of only about ±15% in the O III] flux. Table 1 lists fluxes for a fit using the C IV template profile and arbitrarily assigning about 20% of the O III] flux to another line near λ1665Å. If the other fit were used the He II flux would be increased by 15%, but with no change to the blueward shift of the peak wavelength or to the strong blueward asymmetry of He II.

## 3.4 N III] 1750

The N III] multiplet members at λλ1746.8, 1748.6, 1754.0, 1752.2 and 1749.7Å were fitted in the ratio 0.03:0.21:0.22:0.16:0.38. These are the predicted values from a model computed with the photoionization code *Cloudy* (Ferland 2002) for gas with the $Z = 10$ $Z_{sun}$ composition from Hamann et al. (2002), ionizing photon flux $\log(\Phi) = 19.5$, gas density $\log(n_H) = 9.5$, and all other parameters as in the baseline model used by Korista et al. (1997). In this case a synthetic blend constructed from the N IV] profile clearly gives a much better fit than ones constructed from either the He II or the C IV template profiles. This is illustrated in Figure 5. Nearly identical results were obtained using solar abundances. We conclude that the individual components of N III] λ1750 have profiles which are nearly identical to that of N IV] λ1486.

The observed N III] blend includes Fe II UV 191. This is not included in the synthetic blends shown in Figure 5, but it is well-fitted with a single component at λ1786.7 Å using the C IV template profile.

## 3.5 Other Emission Lines

**Ne VIII λ774.** There appears to be an emission feature near the short-wavelength end of the HST spectrum. The spectrum is badly cut up by Lyα forest absorption and the signal:noise ratio is fairly poor in this region, but the profile seems to be broader than would be expected for either just Ne VIII λλ770.4, 780.3 or just O IV λλ787.2,790.2. Hamann et al. (1998) studied this feature in the spectra of a number of QSOs and concluded that it is most likely Ne VIII with a considerably broader profile than the other emission lines. The poor-quality data for Q0353-383 are consistent with that result, although the measured centroid is displaced towards longer wavelengths by about 3Å in the rest frame, suggesting that some O IV emission might also be contributing. We measured the total flux for this feature, interpolating over obvious absorption lines.

**O VI λ1034.** This feature straddles the junction between the HST and the ground-based data, so there is considerable uncertainty in its exact profile. In addition to the uncertainties in the ground-based flux calibration at the observed wavelength of 3050 Å, the wavelength calibration is also uncertain in this region. However, based on the excellent flux agreement between the HST and ground-based data both at the line peak and in the continuum level to either side of the line, we believe that the integrated flux we measure for this line still must be accurate to perhaps 20%. Lyβ λ1025.7 is also expected to be present at about 1/40 of the strength of Lyα, or contributing 20% of the observed

λ1034 feature. The short-wavelength side of the feature is measured from the HST data, for which the wavelength calibration should be reliable. Profile fitting with the various templates shows no evidence for Lyβ emission, but it could still be present and cut out by overlying absorption. For our present purposes we assign all of the observed flux to O VI.

**Upper limits.** For lines that we do not detect, we find upper limits of $F_{line} < 2 \times 10^{-15}$ erg cm$^{-2}$ s$^{-1}$ in the rest wavelength range λ < 1034Å, and $F_{line} < 10^{-15}$ erg cm$^{-2}$ s$^{-1}$ for 1034 < λ < 1215Å, and $F_{line} < 5 \times 10^{-16}$ erg cm$^{-2}$ s$^{-1}$ for λ > 1215Å. These limits are for unblended lines, and were obtained by determining how much of the C IV template profile could be added to the continuum without producing a noticeable emission line. The limits in the ultraviolet are high because of the combination of poor signal:noise and heavy blanketing by Lyα forest absorption lines, but are sufficient to show that except for the Ne VIII blend, the high-ionization lines in the λλ800–1000Å region all have strengths less than about 25% of O VI λ1034 or C IV λ1549. There may be features at the positions of C III λ977 and N III λ991 with about this intensity. These two lines are important abundance diagnostics (Hamann et al. 2002), but with the existing data it is impossible to be sure that they are present.

## 3.6 Summary of measured line strengths

Table 1, column 2 summarizes the total line strengths (relative to C IV λ1549) measured using the methods described above, while column 3 shows the flux associated with a broad component with the N IV] profile in cases where we can make that separation. Columns 5 and 6 list the corresponding rest-frame equivalent widths. The intensity ratios we find here are in reasonably close agreement with those found by Osmer (1980), with the main difference being the way in which the Lyα+NV blend has been divided up between its components.

Table 1 also lists emission line intensities and equivalent widths for a composite spectrum from Dietrich et al. (2002) for QSOs having the same continuum luminosity as Q0353-383. The composite spectrum is for 162 objects in a 0.5 dex luminosity bin centered on $\log(\lambda L_{1450}) = 46.90$, while for the same cosmological parameters ($H_0 = 65$ km s$^{-1}$ Mpc$^{-1}$, $\Omega_M = 0.3$, $\Omega_\Lambda = 0$) Q0353-383 has $\log(\lambda L_{1450}) = 46.94$, where $L_{1450}$ is the monochromatic continuum luminosity at rest λ1450Å in units of erg s$^{-1}$Å$^{-1}$.

We confirm the previous result of Osmer (1980), that NIV] λ1486 and NIII] λ1750 are considerably stronger in Q0353-383 than in most QSOs. The intensities of N IV] and N III] relative to C IV are 10 and 6 times higher, respectively, in Q0353-383 than in the composite, while the equivalent widths are 5 and 2 times higher. N V λ1240 is also about two times stronger relative to C IV than in the composite spectrum. However, comparison of the equivalent widths shows that C IV, C III] and O III] are much weaker relative to the continuum in the spectrum of Q0353-383 than in that of the composite QSO, N V has the same equivalent width in both spectra, while N IV] and N III] are stronger relative to the continuum. The total Lyα equivalent width is also considerably higher in Q0353-383 than in the composite spectrum, but this is due to the strong narrow spike. The broad Lα component in Q0353-383 is very similar to the total Lyα line in the composite QSO. Figure 1 shows the spectrum of the composite QSO overplotted on that of Q0353-383, and clearly illustrates these effects in the equivalent widths.

# 4.0 Why are the nitrogen lines so strong?

We explore here three possible explanations of the unusual spectrum of Q0353-383. They are: unusual physical conditions (density or ionizing flux) in the BLR clouds, unusual ionizing continuum shape, and unusual chemical abundances.

## 4.1 Normal chemical abundances, but unusual density or ionizing flux?

Could the unusually strong N lines arise from a gas with normal (solar) chemical abundances but peculiar conditions of density or ionizing flux? If the reverberation results from the lower luminosity AGN apply also to luminous quasar broad emission line regions, the BLR extends over a wide range in radius. However, Q0353-383 might be an exception in which most of the BLR gas is located in some small region of the ionizing photon flux ($\Phi$) vs. gas density ($n_H$) plane. We have searched through this parameter space to see if there are any regions that produce greatly enhanced nitrogen lines for any reason. Korista et al. (1997) and Hamann et al. (2002) show contour plots of the equivalent widths and intensity ratios of many of these lines over the log($\Phi$) – log ($n_H$) plane, for solar chemical abundances and a variety of ionizing continuum shapes. We have examined these and similar diagrams, for metallicities up to $Z = 5\ Z_{sun}$, and find that there is no flux-density combination that produces a N IV/C IV intensity ratio larger than about 0.2 at any location where the gas produces these lines with any reasonable efficiency. Similarly, the strong N III] lines cannot be explained in this way. We rule this out as an explanation of the unusual strength of the nitrogen lines.

## 4.2 Unusual ionizing spectrum shape?

Does Q0353-383 produce an ionizing continuum with a highly unusual energy distribution that somehow drives up the strengths of the nitrogen lines? Our models appear to rule this out over the part of the parameter space that has been explored by Korista et al. (1997) and Hamann et al. (2002), but it is still important to look at the actual observations to see if there is any evidence for a continuum shape completely different from those of most QSOs.

Our combined HST and ground-based spectrum covers the rest wavelength range $\lambda\lambda$570-3340 Å. In addition, Dr. Paul Green has generously measured for us from the ROSAT data base an x-ray upper limit of $F(0.1\text{-}2.4\ \text{kev}) < 1.5 \times 10^{-13}$ erg cm$^{-2}$ s$^{-1}$. These data points are shown on Figure 6, where the x-ray upper limit has been converted to $F_\nu$ units assuming that the x-ray energy distribution is a broken power-law of the form $F_\nu \propto \nu^{-1.6}$ below 1 keV, and $F_\nu \propto \nu^{-0.9}$ above 1 keV. The UV-optical data have been corrected for Galactic extinction of E(B-V) = 0.006 mag (Schlegel et al. 1998).

Also shown on Figure 6 is a broken power-law distribution which is a current best guess at the typical QSO energy distribution (Zheng et al. 1997; Laor et al. 1997), and which is used by Hamann et al. in the models to which we will compare our data in Section 4.3. This energy distribution has the form $F_\nu \propto \nu^{-\alpha}$ with $\alpha$ = -0.9, -1.6 and –0.6 in the energy ranges 50 keV –1 keV, 1 keV – 912 Å and 912 Å – 1 µm, respectively, with a steep cutoff beyond 1µm.

On Figure 6, the model energy distribution has been arbitrarily normalized to match the observed continuum level at λ912 Å. The two energy distributions are in reasonable agreement except that the observed HST spectrum dips sharply down at wavelengths below about λ800 Å. This may well be due to Ly-continuum absorption in the Lyα forest, so we conclude there is no evidence to suggest that Q0353-383 has an unusual continuum shape. A deeper X-ray exposure, leading to either a detection or a significantly tighter upper limit, would help to clarify this point. However, there is no evidence so far to indicate that that the continuum energy distribution of Q0353-383 is wildly peculiar.

## 4.3 High metallicity?

The best explanation for the strong nitrogen lines in Q0353-383 is that nitrogen is overabundant by a factor of ~15 relative to solar N/C and N/O ratios. Hamann & Ferland (1993, 1999) have argued previously that the strength of N V λ1240 relative to C IV and He II is a good indicator of the overall metallicity of the BLR. Hamann et al. (2002) showed that N III] and N IV] also get stronger relative to the recombination lines of other elements as the metallicity increases. Nitrogen behaves in this way because for solar abundances and above it is a secondary element, produced mainly in the CNO cycle, so that abundance ratios such as N/O and N/C scale directly with the overall metallicity $Z$. Here we adopt the simple secondary scaling, N/O ∝ O/H, and consider whether the great strengths of the nitrogen lines in Q0353-383 are consistent with an exceptionally high metallicity in this object.

First we compare the total flux in each emission line to models in which the chemical abundances and the ionizing continuum shape were varied. The models are from Hamann et al. (2002), whose results are presented in the form of the intensity ratios of line pairs that are likely to convey information about the metallicity while being relatively insensitive to the ionizing continuum shape. Their figure 4 shows the ratios after integrating each line over an assumed distribution of clouds on the $\log(\Phi) - \log(n_H)$ plane (as in the LOC models of Baldwin et al. 1995). They varied the metallicity $Z/Z_{sun}$ for three different ionizing continuum shapes, including the broken-power-law continuum shape described above in Section 3.

However, an important revision is needed to the results of Hamann et al (2002) and earlier papers. Recent work indicates that the abundances of oxygen, carbon, and iron in the Sun is significantly lower than was assumed in our previous studies (34%, 31%, and 13% smaller respectively; Allende Prieto et al. 2001, 2002, Holweger 2001). The nitrogen abundance is unchanged, so the solar N/C and N/O ratios are now significantly larger. The abundances used by Hamann et al. were scaled from the previous best solar values, taken from Grevesse and Noels (1993), which resulted in a N/O ratio about 35% too large. The impact on the high $Z$ models used by Hamann et al. is that the same N V (relative to other strong lines) now occurs at about 30% lower $Z$ than before. To take this into account, we use Figure 5 from Hamann et al, but subtract 0.11 from the log $Z/Z_{sun}$ values given on the x-axis.

Of the 12 intensity ratios considered by Hamann et al., we can measure in Q0353-383 the following: N III]/C III], N III]/O III], N IV]/O III], N IV]/C IV, N V/He II, N V/C IV,

N V/O VI and N V/(C IV+O VI). We have converted these to metallicities using the Hamann et al. results for the broken-power-law continuum shape, since it is in the best agreement with the limited data on the continuum of Q0353-383. The results are listed in column 3 of Table 2. Figure 7 plots our measurements on top of the Hamann et al. results. The ratios involving N V are calculated using both the maximum and minimum strength for this line, and are shown on the figure as a range linking together the two resulting values. In all cases the measured ratios indicate $\log(Z/Z_{sun}) > 0.8$. We averaged together the derived metallicities, but for line ratios involving N V we used only the mean values for the metallicities found from N V/He II and N V/(C IV+O VI). The result is $\log(Z/Z_{sun}) = 1.2$.

The observed equivalent widths of these lines (Table 1) also generally behave as is expected for the case of high metallicity. Since the N abundance has increased greatly at the expense of C and O, the N lines take over a greater share of the cooling and become stronger relative to Ly$\alpha$, while the C and O lines carry less cooling and become weaker. This expected behavior was illustrated in Ferland et al. (1996, their Fig. 15) for a model for a single BELR cloud. As a further check, Table 3 compares observed line intensities relative to Ly$\alpha$, which again shows the relative amount of cooling provided by each line, to the predicted intensity ratios from our LOC models for solar and 15× solar metallicity. This approach is much more model dependent than the intensity ratios used in Table 2, because Ly$\alpha$ is formed over a wide range on the $\log(\Phi) - \log(n_H)$ plane, so the calculated intensity ratios depend more heavily on the assumed distribution of gas over this plane (here we have used the standard LOC parameters from Baldwin et al. 1995). We see from Table 3 that the agreement is not perfect, and that clearly the LOC models predicts both N III] and C III] to be stronger than observed. However, in general (with the exception of C III]) the N lines become stronger relative to Ly$\alpha$ while the C and O lines become weaker. In summary, the integrated strengths of the individual emission lines support the conclusion that C and O have been converted to N by secondary enrichment.

## 4.4 Chemical Enrichment Anomalies?

It is important to note that the large nitrogen enhancement in Q0353-383 cannot result from "pollution" by the ejecta of one or a few nearby stars. The reason is that the mass of BLR in luminous quasars like Q0353-383 is at least $10^3$ M$_{sun}$, and therefore a substantial stellar population with mass $\geq 10^4$ M$_{sun}$ must be involved in the enrichment (Baldwin et al. 2003). The minimum mass of the enriching stellar population could actually be several orders of magnitude larger if the super-massive black hole, accretion disk and BLR were all formed from gas with the same chemical history. Observations of old, metal-rich stars in galactic nuclei support the idea that substantial, bulge-size stellar populations participate in the prompt enrichment observed in QSOs (eg. Friaca & Terlevich 1998, Hamann & Ferland 1999, Romano et al. 2002, and references therein).

Another possibility is that the large nitrogen enhancement in Q0353-383 is caused by a rare, extreme departure from the nominal N/O $\propto$ O/H scaling. Observed enhancements like this have been attributed to the timing soon after a major starburst, where intermediate mass stars are dumping substantial nitrogen into the system while the oxygen enrichment by massive stars has slowed or stopped (Garnett et al. 1990, Henry et al. 2000). However, even in this context, the extreme N/O and N/C enhancements observed in Q0353-383 can be expected only at metallicities well above solar. In particular, positive departures of N/O from the nominal N/O ∝ O/H curve (passing through solar abundance ratios) should be less than a factor of ~2 in the metal-rich regime. Therefore, these considerations still suggest that the metallicity in Q0353-383 is at least 5 – 10 $Z_{sun}$.

## 5.0 Conclusions

One important goal of this paper has been to test whether or not the strengths of the nitrogen lines, particularly of N V λ1240 relative to C IV λ1549 and He II λ1640, is a valid abundance indicator in AGN spectra. We find that the best explanation of the unusually strong nitrogen lines in Q0353-383 is that nitrogen is overabundant. In this particular case of extreme metal enrichment, the strengths of NV, N IV] and N III] relative to lines of C, O and He all indicate very high values of the metallicity $Z$ using the Hamann et al. (2002) technique. This indicates that the N V line is indeed a valid abundance indicator, and gives considerable support to our technique of using N V measurements in the emission line spectra of more typical quasars to trace the general trend of chemical evolution in the centers of massive galaxies at large lookback times.

Taking into account the revised solar abundances mentioned in §4.3, our analysis points to an overall metallicity $Z \sim 15\ Z_{sun}$. The method used in §4.3 really measures the N/O and N/C abundance ratios. The results are converted to the overall metallicity $Z$ through the assumption that the N is a secondary element, so that N/O and N/C scale directly with $Z$ (see Hamann et al. 2002 for details). Our approach to converting emission line intensities into abundances is basically a simple matter of energy balance – the photoionized plasma cools by exciting the observed lines – the fact that the N lines are so strong reflects the fact that this coolant has a very high abundance relative to either C or O. The numerical simulations allow us to quantify how abundant N must be. Conservation of energy underlies our approach, which does not depend on detailed measurement of electron temperature or density.

However, there still are significant uncertainties in this derived metallicity. The 30% rescaling due to the new solar abundances points to one (possibly surprising) area of uncertainty – abundance determinations are difficult even for an object as simple as the Sun. The scatter in the horizontal direction on Figure 7 indicates another: the different metallicity indicators give results ranging from 6 – 25 times solar, showing that the combination of the BLR models and the sets of abundance used by Hamann et al. do not perfectly describe Q0353-383. Nuclear yields at high $Z$ are another question.

A further point to consider is that the N/O ∝ $Z$ relation probably is not strictly followed at every moment in a galaxy's history. Large-scale starbursts may temporarily move a galaxy to one side or the other of the mean relation. For example, Coziol et al. (1999) noted that H II regions in starburst nucleus galaxies generally do follow the expected N/O ∝ $Z$ scaling at higher metallicities (where secondary N production dominates), but that there is significant scatter in this relationship (factors of 2–3 at any given $Z$). They suggested that the scatter is caused by the occasional timed release of nitrogen by large numbers of stars formed in specific starburst events. Overall, however, large samples of AGNs should still roughly follow the average N/O ∝ $Z$ scaling found in galaxies (for $Z >$ 0.2 $Z_{sun}$ where secondary N production dominates, Hamann et al. 2002, Hamann & Ferland 1999, Villa-Costas & Edmunds 1993, Van Zee et al. 1998).

So what was going on in Q0353-383, 11 Gyr ago? The mass of metal-rich gas needed to explain the observed emission line luminosities in Q0353-383 must be large. If we are to believe, from reverberation results for lower-luminosity AGN, that the BLR in high-luminosity objects such as Q0353-383 is distributed over a factor of 20 – 30 in radius, then much of the BLR gas must emit at very low efficiency per unit mass. Baldwin et al. (2002) found that a reasonable estimate of the total BLR mass for such a distributed system is $M_{BLR} \sim 10^3–10^4$ ($L_{1450}/10^{44}$ erg cm$^{-2}$ s$^{-1}$ Å$^{-1}$)(EW(Lyα)/56Å) $M_{sun}$. This estimate has been scaled from the Lyα luminosity to $L_{1450}$ by using a typical Lyα rest equivalent width of EW(Lyα) = 56Å. For Q0353-383, EW(Lyα) = 87Å and $L_{1450}$ = 6.0×10$^{43}$ erg s$^{-1}$ Å$^{-1}$, with the larger EW(Lyα) balancing out the lower $L_{1450}$ so that the same BLR mass estimate will hold.

The simplest explanation for the N-rich gas commonly found in the BLRs of luminous quasars is that it is metal-rich gas falling in from the surrounding protgalaxy on its way to fuel the central engine. As is discussed by Baldwin et al (2003), the high metallicity of the BLR gas, particularly in Q0353-383, shows that this material must have come from a stellar system with at least an order of magnitude more mass. For example, in the Hamann & Ferland (1993) models, metallicities of ~10$Z_{sun}$ are reached at a time when the gas fraction has dropped to ~3%, meaning that there must be at least 30 times more material ( ~10$^5$ M$_{sun}$) locked up in stars. However, as is discussed by Baldwin et al. (2002), there is likely to be far more gas present that is associated with the BLR but does not produce UV-optical emission lines, and the 10$^8$–10$^9$ M$_{sun}$ of material that has already made its way into the central black hole is likely to also have left behind some correspondingly larger mass in stars in the surrounding galaxy. Thus we are sampling a phenomenon covering a significant part of the central region of the host (proto)galaxy.

Our interpretation of Q0353-383 is that we observe it at the very end of the epoch of rapid metal enrichment that we expect to occur in the central parts of galaxies. We expect rapid enrichment of the ISM in the central regions of the host galaxies for the simple reason that the cores of massive galaxies have the highest densities. Hamann & Ferland (1999; see their Fig 13) show that the final state of chemical evolution in such an environment should in fact lead to $Z \sim 10 - 20\ Z_{\odot}$, the sort of value found in Q0353-383, and plateau at this level after ~1 Gyr.

Q0353-383 has a highly unusual spectrum; we have seen only two possibly similar cases of such strong N IV] and N III] among many hundreds of quasar spectra that we have studied. The scarcity of Q0353-383-like objects may simply reflect the amount of time a quasar remains active once the stellar cluster has become fully chemically mature. In most cases the quasar activity must have died out, presumably from exhaustion of the gas supply within capture range of the central black hole, by the time such high metallicities are reached. In fact, the Hamann & Ferland (1993) models terminate at $Z \sim 10–20\ Z_{sun}$ *because* of gas depletion (defined as the gas fraction dropping to 3%).

We can use a similar technique to estimate when QSOs came to life. Photoionization simulations show that low $Z$ gas still produces bright emission lines, so no selection effects should make low $Z$ objects hard to discover. No low metallicity quasars are observed, suggesting that the quasar does not become luminous until the chemical enrichment is well progressed. The Hamann & Ferland models indicate that solar metallicity should be reached at the center of a massive galaxy after ~ 0.2 Gyr. Perhaps this is the timescale needed to build a massive black hole in the galaxy's core, presumably by coalescence of stellar remnants. Therefore, the range of chemical abundances suggests that active QSOs occur in massive host galaxies during the period when they are between 0.2 Gyr and 1Gyr old. The typical lifetime of an individual QSO can therefore be at most ~0.8 Gyr, although it also could be much shorter if the central engine is only fed for a short time somewhere within this period. This can be more carefully constrained by studying the statistics of BLR metallicities at high redshift.

The few cases of extremely high metallicity have also turned out to be among the most luminous quasars. The metallicity-luminosity correlation in QSOs (Hamann & Ferland 1993) could be interpreted as an age-metallicity-luminosity correlation. If individual QSOs do in fact stay active (perhaps episodically) over the full 0.8 Gyr span, then the more mature objects would also be more metal rich. The standard accretion-disk theory says that the luminosity depends on both the accretion rate and mass of the central black hole. The metallicity-luminosity correlation would then provide an independent measure of the history of the black hole growth, not tied to using the redshift as the time scale.

We thank the referee, Pat Osmer, for valuable suggestions, and Paul Green for providing the x-ray upper limit. We gratefully acknowledge NASA grant HST-GO-08283 for support of this research project. In addition, FH acknowledges financial support from the National Science Foundation through grant AST 99-84040, and GJF thanks the NSF (AST 0071180) and NASA (NAG5-8212 and NAG5-12020) for support.

| | $I$(line)/$I$(CIV λ1549) | | | Rest Equivalent Width (Å) | | |
|---|---|---|---|---|---|---|
| Line | Q0353-383 total[1] | Q0353-383 broad[1] | Composite QSO[2] | Q0353-383 total | Q0353-383 broad | Composite QSO[2] |
| Ne VIII λ774 | 0.96 | | | 16 | | |
| O VI λ1035 | 0.92 | | 0.34 | 10 | | 11 |
| Lyα | 9.0-7.0 | 8.8 | 2.0 | 111-87 | 57 | 61 |
| N V λ1240 | 1.5-2.5 | 4.3 | 0.98 | 19-31 | 28 | 29 |
| O I (+ Si II?) λ1305 | 0.14 | | 0.10 | 2 | | 3 |
| O IV+Si IV λ1400 | 0.40 | | 0.26 | 5 | | 9 |
| N IV] λ1486 | 0.30 | 0.57 | 0.03 | 5 | 5 | 1 |
| C IV λ1549 | 1.00 | 1.00 | 1.00 | 17 | 9 | 43 |
| He II λ1640 | 0.20 | | 0.16 | 4 | | 8 |
| O III] λ1663 | 0.10 | 0.10 | 0.12 | 2 | 1 | 6 |
| N III] λ1750 | 0.47 | 0.88 | 0.08 | 9 | 9 | 4 |
| Fe II uv191 λ1787 | 0.05 | | | 1 | | |
| Al III λ1857 | < 0.05 | | 0.11 | < 1 | | 6 |
| Si III] λ1892 | 0.05 | | 0.02 | 1 | | 1 |
| C III] λ1909 | 0.38 | 0.38 | 0.27 | 9 | 5 | 17 |
| Mg II λ2798 | 0.34 | 0.34 | 0.31 | 12 | 6 | 37 |

**Table 1**

**Measured Intensities and Equivalent Widths**

[1] Normalized to F(CIV) = $9.8\times10^{-15}$ erg cm$^{-2}$ s$^{-1}$ for the total line fluxes, and $5.1\times10^{-15}$ erg cm$^{-2}$ s$^{-1}$ for the broad components.

[2] Values for log$\lambda L_\lambda$ = 46.90 composite QSO spectrum from Dietrich et al. (2002).

| Table 2. Metallicities derived from observed line strengths. | | |
|---|---|---|
| Lines | Observed log line ratio | log $Z/Z_{sun}$ |
| N III]/C III] | 0.09 | 0.94 |
| N III]/O III] | 0.67 | 1.28 |
| N IV]/O III] | 0.48 | 1.41 |
| N IV]/C IV | -0.52 | 0.79 |
| N V/He II | 0.86 – 1.10 | 1.13 –1.39 |
| N V/C IV | 0.17 – 0.41 | 1.19 –1.54 |
| N V/O VI | 0.21 – 0.45 | 1.08 –1.39 |
| N V/(C IV+O VI) | -0.11 – 0.13 | 1.13 –1.48 |

| Table 3 Line Intensities Relative to Ly$\alpha$ = 100 | | | |
|---|---|---|---|
| Line | Observed | LOC Models | |
| | Q0353-388 | $Z = Z_{sun}$ | $Z = 15\ Z_{sun}$ |
| O VI $\lambda$1034 | 10 – 12 | 23 | 19 |
| Ly$\alpha$ | 100 | 100 | 100 |
| N V $\lambda$1240 | 17 – 33 | 5 | 37 |
| N IV] $\lambda$1486 | 3 – 4 | 2 | 19 |
| C IV $\lambda$1549 | 11 – 13 | 43 | 24 |
| He II $\lambda$1640 | 2 – 3 | 7 | 4 |
| N III] $\lambda$1749 | 5 – 6 | 1 | 40 |
| C III] $\lambda$1909 | 4 – 5 | 8 | 21 |

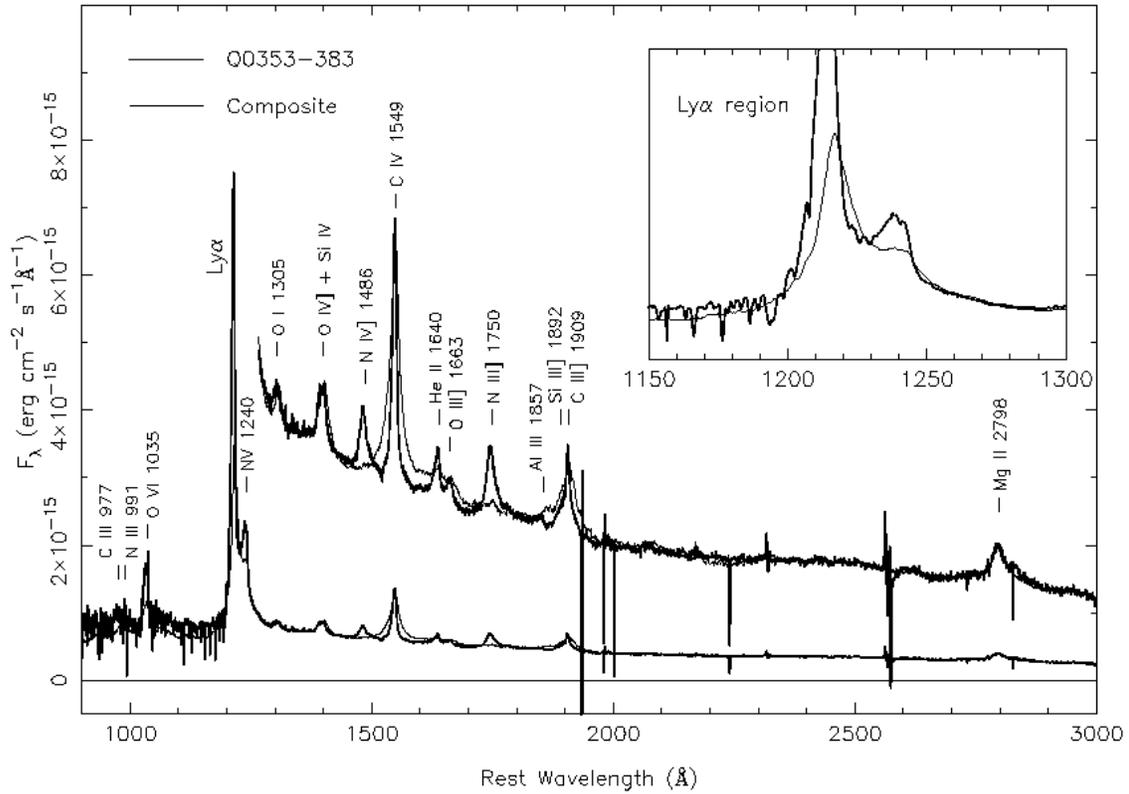

**Figure 1.** Heavy line: Combined HST and ground-based spectrum. The region to the red of Lyα is also shown with the flux scale blown up by a factor of 5. Light line: The $\log(\lambda L_{\lambda-1450}) = 46.90$ composite spectrum from Dietrich et al. (2002). The objects in the composite spectrum lie within a 0.5 dex luminosity bin centered on nearly the continuum luminosity of Q0353-383. The continuum level of the composite spectrum has been normalized to that of Q0353-383 at 1450Å.

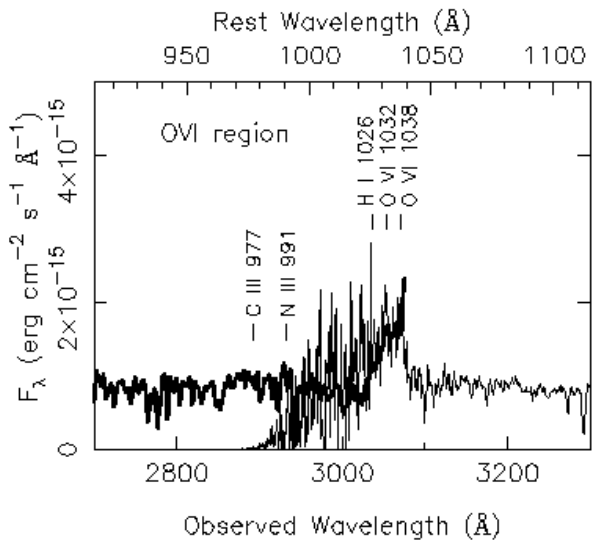

**Figure 2**. The wavelength region including the overlap between the HST spectrum (heavy line) and blue ground-based spectrum (light line), showing the good photometric agreement across the OVI emission line.

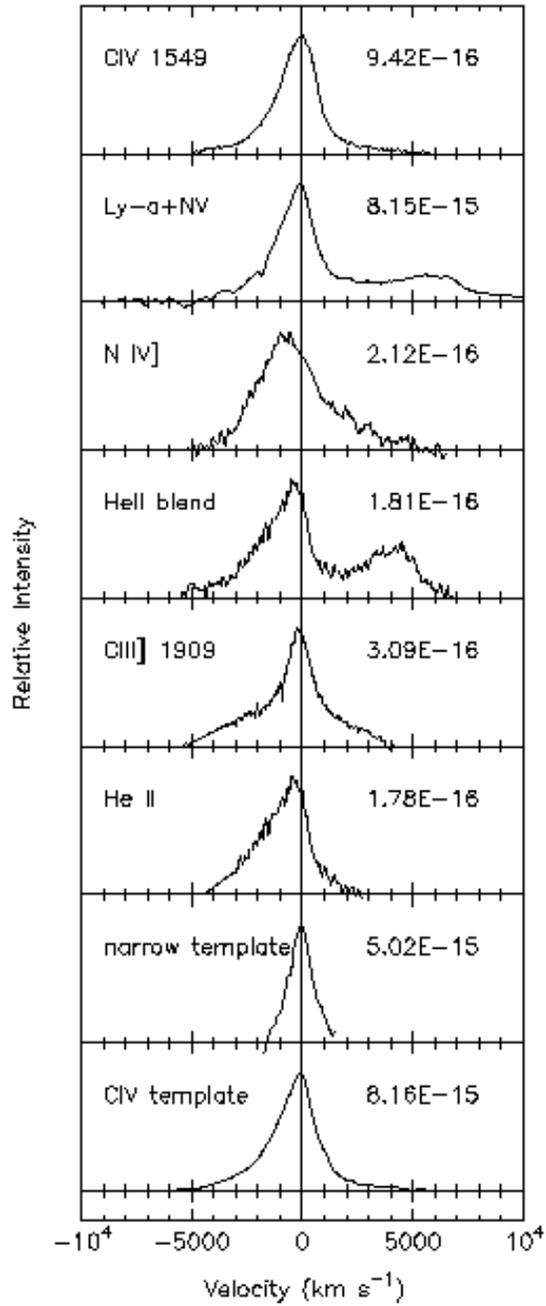

**Figure 3**. Selected continuum-subtracted emission line profiles on the same velocity scale and normalized to have the same peak height. The number on the right side of each frame shows the peak continuum-subtracted $F_\lambda$ in the line.

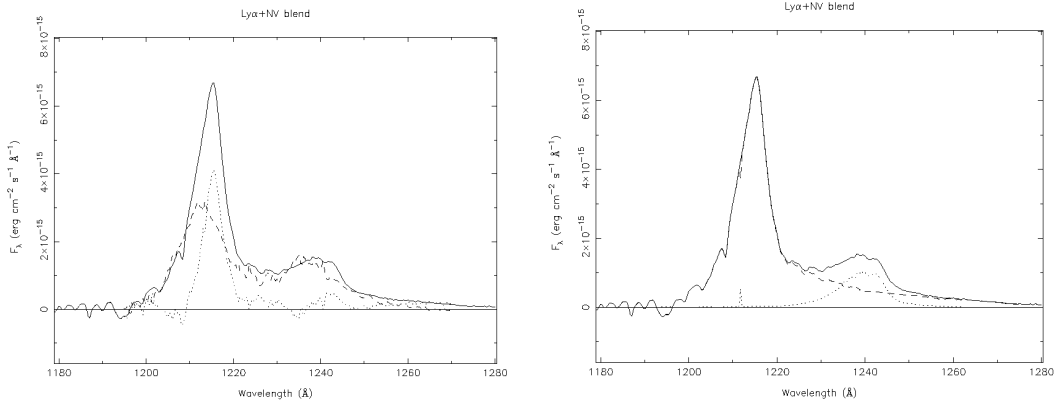

**Figure 4.** (a) The separation of the Lyα-NV blend using the combination of the NIV] template profile and a narrow spike. Solid line: the observed profile. Dashed line: synthetic blend made up of NIV] template profiles shifted to $\lambda\lambda 1215.7$, 1238.8 and 1242.8 Å and accounting for, respectively, fractions 0.45, 0.117 and 0.103 of the total flux in the observed blend. Dotted line: the residual from the fit. The narrow residual spike between $\lambda\lambda 1210$ –1220 Å was then used as our narrow template profile. (b) Solid line: the same observed profile as in Fig. 4a. Dotted line: the NV fit that gives a lower flux limit to the NV flux, as described in the text. Dashed line: the residual after subtracting the NV fit. This gives our upper limit on the Lyα flux, which includes the strong red wing.

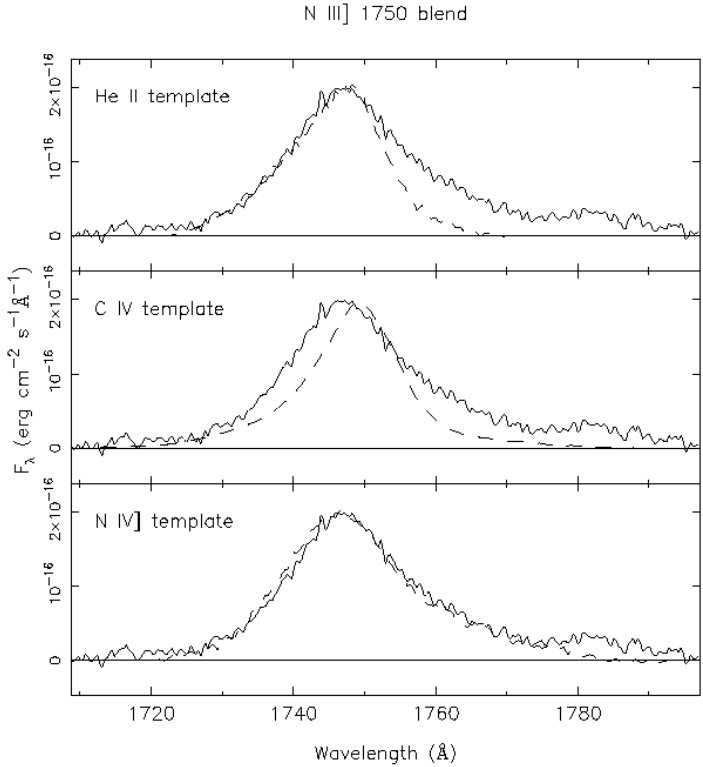

**Figure 5.** Three attempts at fitting the observed N III] $\lambda 1750$ blend. The template profiles are listed in each panel. The N IV] template obviously provides by far the best fit. The weak residual feature at 1787Å is Fe II UV 191.

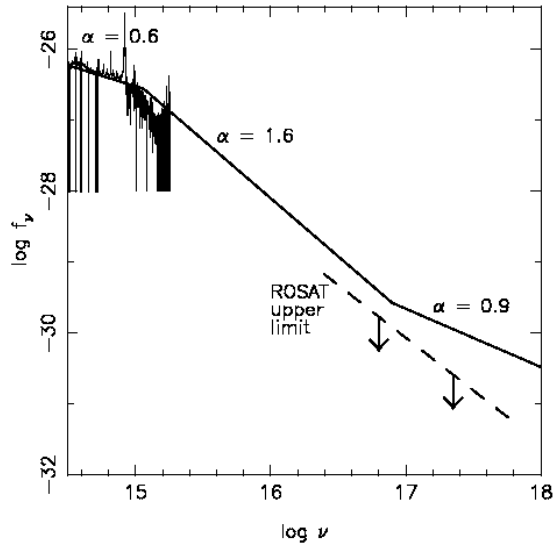

**Figure 6.** The observed continuum energy distribution for Q0353-383, compared to the broken power law representing typical QSOs (Hamann et al. 2002). The lower line shows the ROSAT upper limit.

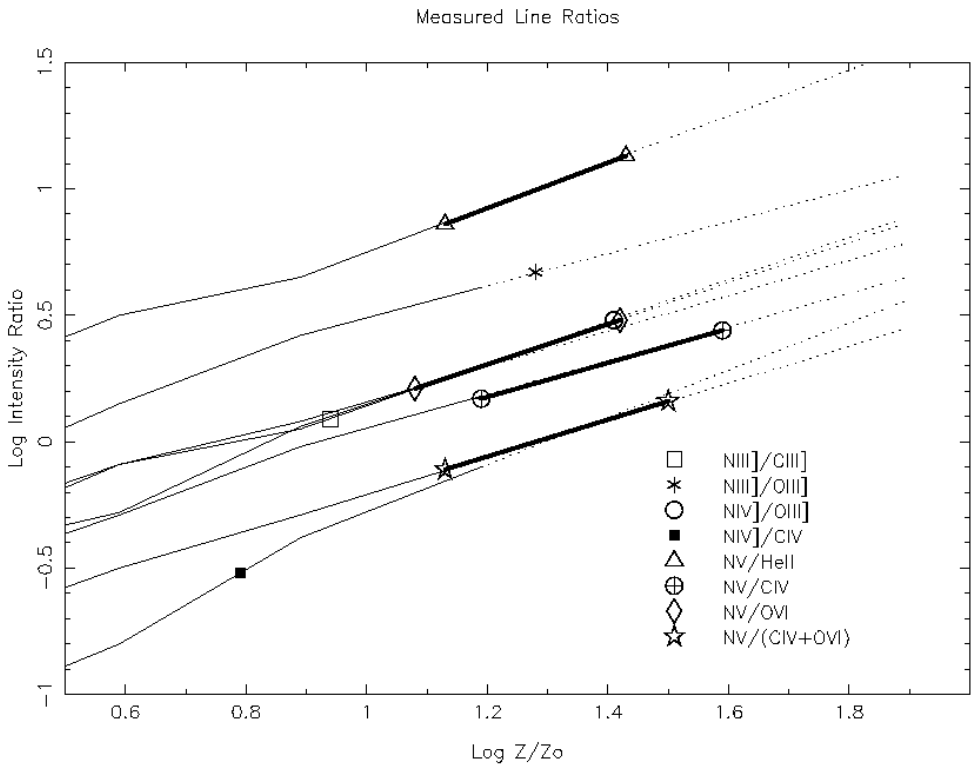

**Figure 7.** Metallicity determination using the Hamann et al models. The model results for the different intensity ratios are represented by the separate lines on the plot, which become dotted lines at metallicites that are an extrapolation beyond the model grid. The symbols show where the measured intensity ratios (see key) fall on the corresponding lines on the plot. The heavy lines connecting pairs of the same symbol show the metallicity range due to the two ways of measuring the NV strength.

21